\documentclass[twocolumn,floatfix%
 reprint, 
 amsmath,amssymb,
 aps, prb
]{revtex4-2}

\usepackage{graphicx}
\usepackage{dcolumn}
\usepackage{bm}
\usepackage{braket}
\usepackage{mathtools}
\usepackage{booktabs}
\usepackage{verbatim}
\usepackage{booktabs}
\usepackage{siunitx}
\DeclarePairedDelimiter\abs{\lvert}{\rvert}%
\usepackage[T1]{fontenc}
\usepackage[utf8]{inputenc}

\begin{document}

\preprint{\LaTeX}

\title{\textbf{Enhancing Neural Network Backflow}}

\author{Kieran Loehr}
 \email{Contact author: kloehr@illinois.edu}
\author{Bryan K. Clark}%
\affiliation{%
  The Anthony J. Leggett Institute for Condensed Matter Theory and IQUIST and NCSA center for Artificial Intelligence Innovation and Department of Physics, University of Illinois at Urbana-Champaign, IL 61801, USA
}%

\date{\today}

\begin{abstract}
Accurately describing the ground state of strongly correlated systems is essential for understanding their emergent properties. Neural Network Backflow (NNBF) is a powerful variational ansatz that enhances mean-field wave functions by introducing configuration-dependent modifications to single-particle orbitals. 
Although NNBF is theoretically universal in the limit of large networks, we find that practical gains saturate with increasing network size. 
Instead, significant improvements can be achieved by using a multi-determinant ansatz. We explore efficient ways to generate these multi-determinant expansions without increasing the number of variational parameters. In particular, we study single-step Lanczos and symmetry projection techniques, benchmarking their performance against diffusion Monte Carlo and NNBF applied to alternative mean fields. 
Benchmarking on a doped periodic square Hubbard model near optimal doping, we find that 
a Lanczos step, diffusion Monte Carlo, and projection onto a symmetry sector all give similar improvements achieving state-of-the-art energies at minimal cost. By further
optimizing the projected symmetrized states directly, we gain significantly in energy. Using this technique we report the lowest variational energies for this Hamiltonian on $4\times 16$ and $4 \times 8$ lattices as well as accurate variance extrapolated energies. We also show the evolution of spin, charge, and pair correlation functions as the quality of the variational ansatz improves.  
\end{abstract}

\maketitle


\section{\label{sec:introduction}Introduction}
Accurate ground state wave-functions of strongly correlated systems are critical to extracting their properties. This is especially true as energy differences between qualitatively different phases of matter can be small. Representing the state exactly is hindered by the exponentially large Hilbert space motivating the use of variational wave-functions.  

Recently, the class of neural network quantum states (NQS) \cite{OriginalNQS}  which uses machine learning architectures as part of the variational wave-function has shown potential for representing quantum many-body states with significant progress on using NQS for spin systems \cite{VariationalBenchmarks, MinSR, SimpleAlgebra_MinSR, KagomeNQS, SpinEx_40, SpinEx_42, SpinEx_43}. Fermion NQS are less well developed with some approaches directly mapping second-quantized configuration to amplitudes \cite{Choo2020_FermionNQS} or modifying the Jastrow factor  \cite{PP_RBM}. Among the most accurate fermion variational wave-functions are the Neural Network Backflow Wavefunctions (NNBF) and low-rank variants such as the hidden fermion approaches which use ML to directly modify the mean-field determinant  \cite{OriginalNNBF,OriginalHFDS,ZejunHFDS,Di_Moire,FermiNet,PauliNet, Andrew_NNBF}. Despite promise, there is still significant room for improvement in these approaches. Various attempts at building on the standard NNBF form have included  using ML architectures other then multi-layer perceptrons (MLP) \cite{TransformerBF, symHFDS, Di_CNN} and mean-fields beyond the Slater determinant including BDG \cite{OriginalNNBF} as well as pfaffians \cite{HFPS}.

In this work, we fix the neural network architecture to an MLP and consider various ways to systematically improve MLP-NNBF.  We focus our tests on the $4 \times 16$ Hubbard model in periodic boundary conditions at $U=8$ and hole doped at $\delta =1/8$ as the paradigmatic example of a physically interesting albeit challenging strongly correlated fermionic system which has been historically used as a benchmark for such methods.  We focus on both the accuracy as well as computational efficiency of various improvements. While MLP-NNBF is universal in the limit of enough neurons \cite{OriginalNNBF} in practice we find that the benefits of increasing neurons begins to saturate when the cost to evaluate the network becomes approximately equal to the cost of evaluating the determinant(s).  

Instead, we find that one can gain significantly by using a sum of multi-determinant MLP-NNBF's \cite{FermiNet, PauliNet}. While formally, one could simply independently optimize all these determinants, in practice this is often difficult and time-consuming. Instead, we primarily focus on ways of generating these determinants without increasing the number of parameters.  We do this either by using Lanczos steps \cite{VMC-Lanczos, Lanczos-Example, TBF-Lanczos} or symmetry projection \cite{GustavoSym,VMC_Sym_Ex79,VMC_Sym_Ex81,VMC_Sym_Ex82} both of which generate a multi-determinant expansion from a single MLP-NNBF.  First optimizing the single MLP-NNBF and then applying either Lanczos or symmetry projection gives similar but significant benefits resulting in the lowest energies available for this Hamiltonian; this state-of-the-art calculation took roughly 200 GPU hours. By further optimizing the projected states directly, we gain an additional approximately 45\% towards the variance extrapolated ground state at a cost of almost 3000 GPU hours.  

We additionally investigate the promotion of the determinant mean field to that of a pfaffian finding that the energy improvement is comparable to increasing neurons, but with a larger computational cost. We also apply diffusion Monte Carlo (DMC) \cite{DMC_Review} to the single MLP-NNBF, and find that it can frequently give energies comparable to Lanczos and unoptimized symmetry projection.  Finally, we compute spin, charge, and pairing correlations and illustrate how they evolve under increasingly accurate variational wave-functions.  

\section{\label{sec:methods}Methods}

\subsection{\label{subsec:nnbfansatz}NNBF Ansatz}
The NNBF ansatz extends a mean-field wavefunction by introducing configuration-dependant backflow corrections that can alter both the amplitude and sign structure of the state \cite{OriginalNNBF}.  The amplitude for a configuration $\mathbf{n}$ is
\begin{equation}
    \psi_{NNBF}(\textbf{n}) = \langle{\textbf{n}}|{\psi_{NNBF}}\rangle = \det(\textbf{u}(\textbf{n}) [\textbf{n}])
\end{equation}
where $\textbf{u}(\textbf{n})$ is the output, an $N\times L$ matrix,   of an MLP with input $\textbf{n}$. $[\textbf{n}]$ notates selecting the columns of $\textbf{u}(\textbf{n})$ corresponding to the occupied sites resulting in a $N \times N$ matrix. Here both spin sectors are combined into a single matrix (e.g. $N$ is the sum of the spin up and down electrons and $L$ is twice the number of physical sites), which offers greater representational ability than the product of separate spin matrices \cite{LinLinFullDet}. 

Our MLP is fully connected with two hidden layers each followed by a layer normalization and relu activation. The input layer is set by $\ket{\mathbf{n}} = \ket{n_0,n_1,...,n_L}$ where $n_i \in \{-1,1\}$ denotes an empty or occupied spin orbital.  The last output layer contains a bias but no non-linear function; this bias can be interpreted as a fixed mean-field solution onto which the neural network provides additive configuration-dependent backflow corrections. The accuracy of this MLP-NNBF can be systematically improved by increasing the number of hidden neurons, $n_h$.

\subsection{\label{subsec:optimization}Optimization}

To enhance training efficiency and mitigate convergence to local minima, we update our network as 
\begin{equation}
    \theta_{i+1} = \theta_i - \textbf{s}_i\delta_{L}\times\text{sign}[\nabla_\theta E_\theta]
\end{equation}
where $\theta_i$ denotes the parameter vector at iteration $i$, $\delta_{L}$ is the learning rate, and $\textbf{s}_i$ is a vector of scalars drawn uniformly from the interval $[0,1)$ at each iteration \cite{randSGD_Clark,randSGD_Sandvik,randSGD-Neuscamman,signSGD}. The deviation of this update step from the gradient helps prevent the optimizer from becoming trapped in shallow energy basins.

To accelerate convergence, we first optimize the mean-field Slater determinant before introducing the backflow corrections. During the final stages of training, we apply an exponential learning-rate decay to allow the optimizer to refine the variational minimum by flowing to the minima of narrow funnels. Further optimization details are available in Appendix \ref{sec:OptimizationDetails}.  

All optimizations reported in this paper were performed using H100 GPUs on GH200 superchips. Most optimizations were performed using a single H100 GPU, but longer optimizations were parallelized across 4 GPUs. GPU times are reported as cumulative GPU usage (e.g. parallelized times are 4 times the wall clock time). All optimization times include the cumulative cost of optimization for a given point which includes (where specified) optimizing smaller ansatz first. 

\section{\label{sec:results}Results}

To assess the improvements to the NNBF ansatz, we benchmark on the two-dimensional Fermi-Hubbard model on a square lattice
\begin{equation}
    \hat{\mathcal{H}} = -t \sum_{\langle ij \rangle, \sigma} (\hat{c}_{i,\sigma}^\dagger \hat{c}_{j,\sigma} + h.c.) + U \sum_i\hat{n}_{i,\uparrow}\hat{n}_{j,\downarrow}\label{eq:hubb_hamiltonian}.
\end{equation}
We focus on the computationally challenging regime of strong repulsive interactions $U = 8$, $t=1$ and small hole doping $\delta = 1/8$ in periodic boundary conditions. We primarily report results on a $4\times16$ lattice outside of section \ref{subsec:4x8Lattice} where we report on the $4\times 8$ system. To compare between different lattice sizes more easily, we always report the energy per site and its variance throughout this work.
\begin{figure}[htbp]
    \centering
    \includegraphics[width=0.48\textwidth]{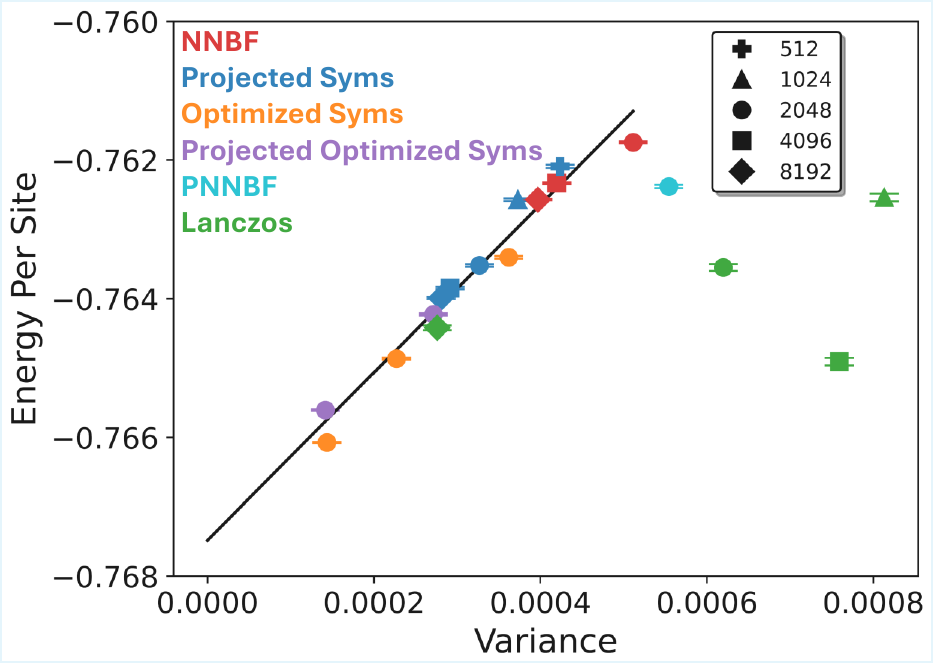}
    \caption{Energy vs. variance of different variational wave-functions  with a  linear extrapolation to zero variance taken through the base NNBF and all symmetrized versions (e.g. red, blue, orange and purple points) whose energy was less then -0.7615. Symbols denote $n_h$.}
    \label{fig:VarianceExtrapolation}
\end{figure}

\begin{figure}[htbp]
    \centering
    \includegraphics[width=0.48\textwidth]{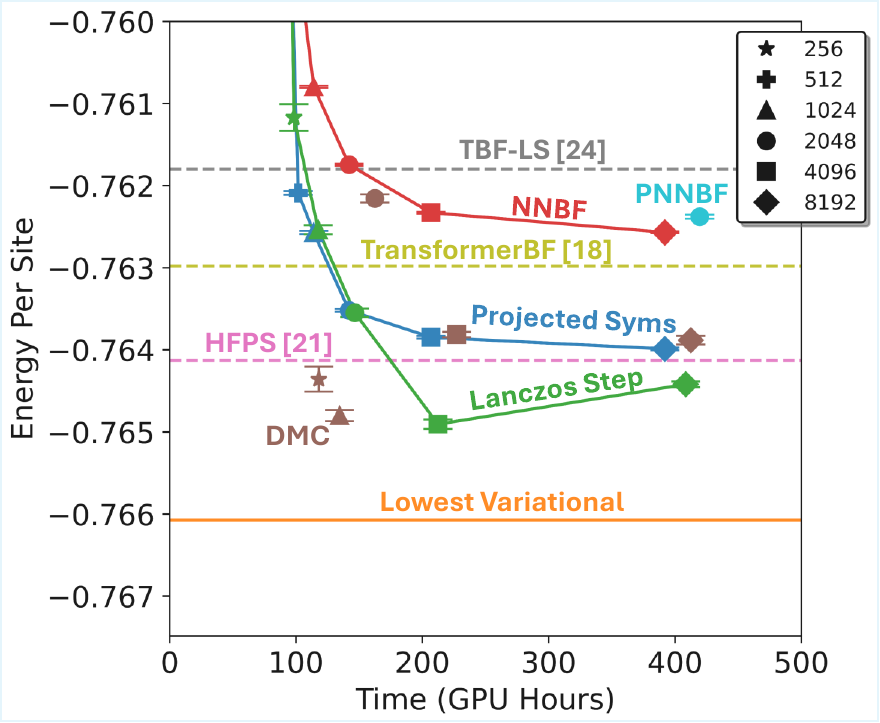}
    \caption{Variational energy of NNBF approaches as a function of GPU training hours. Symbols denotes $n_h$. Comparison with other works shown as dotted lines.  The lowest variational energy achieved in this work is shown as a solid line. The Lanczos step optimization and DMC for $n_h = 512$  failed and are excluded.}
    \label{fig:NeuronEnhancements}
\end{figure}

\begin{figure}[htbp]
    \centering
    \includegraphics[width=0.48\textwidth]{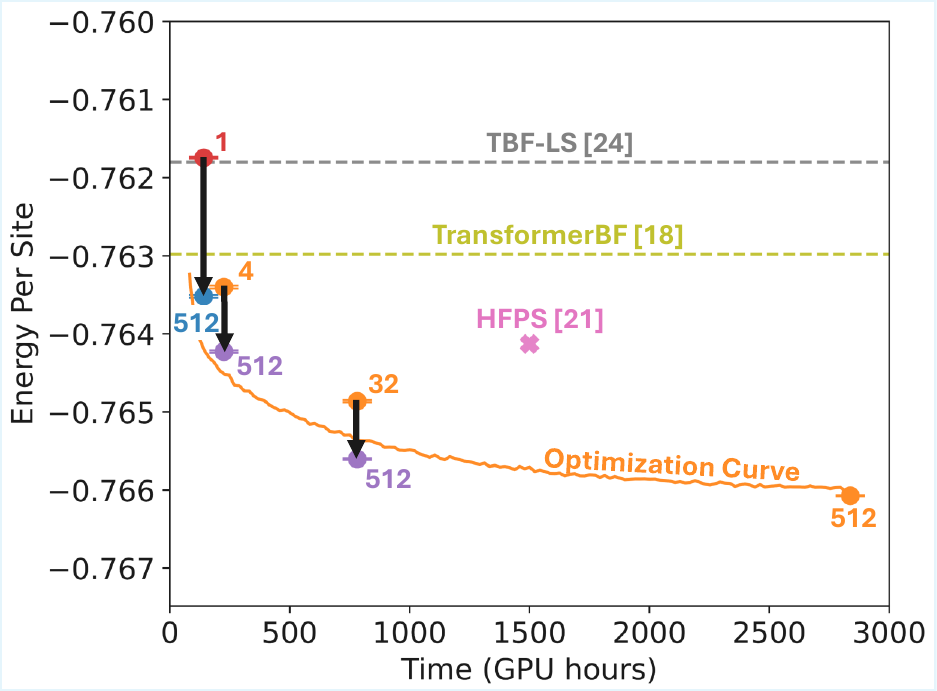}
    \caption{Variational energy versus GPU training hours for different symmetric operations at $n_h=2048$. We consider both optimization with a fixed number of symmetries (orange), and projection from these points to the full symmetry sector (purple). The numbers next to each point correspond to the number of symmetries. The projection from the non-symmetrized (red point) to full symmetries (blue point) is also seen in Fig.~\ref{fig:NeuronEnhancements}.  Comparison with other works shown as dotted lines and a pink cross are all higher in energy then the projected optimized symmetries. The optimization curve is the optimization for the optimized fully symmetric state.}
    \label{fig:SymmetryOptimization}
\end{figure}

\noindent \textbf{\textit{Variance Extrapolation:}} 
Throughout this manuscript, we consider a number of NNBF enhancements measuring both the energy and their variance (see Fig.~\ref{fig:VarianceExtrapolation}). Close to the true ground state, the energy and variance should be linearly related and, in practice, this can be true even outside this regime if one chooses a uniformly related set of wave-functions. Here we do a linear extrapolation through all sufficiently low energy points of our standard NNBF models and all forms of symmetry projections (discussed in detail in later sections). We expect the extrapolated energy, $-0.76748$, which is not necessarily variational, to be a good proxy for the true ground state energy of our Hamiltonian and use this value as the ``ground truth'' to report fractional improvements of our different methods discussed in this paper.  The y-axes of Fig.~\ref{fig:NeuronEnhancements}, Fig.~\ref{fig:SymmetryOptimization}, and Fig.~\ref{fig:BaseNeurons} use this value as its minima. 

\noindent \textbf{\textit{Increasing neurons:}} With 64 hidden neurons and a single-layer MLP, NNBF originally reached energies of $-0.746$ per site \cite{OriginalNNBF}. 
Both significantly increasing neuron width, the number of layers, and using full determinants instead of products of spin determinants yields much more accurate results \cite{LinLinFullDet, ZejunHFDS}. 
In our implementation, a two-layer MLP with 8192 neurons reaches an energy of $-0.76258(1)$ within 400 GPU hours. This energy is lower than recent results using a tensor backflow ansatz with a Lanczos step \cite{TBF-Lanczos}.

Fig.~\ref{fig:BaseNeurons} and Fig.~\ref{fig:NeuronEnhancements} show the dependence of the variational energy on the number of hidden neurons, $n_h$ for standard NNBF. The energy decreases with neurons at small $n_h$ but improves much more slowly as $n_h$ increases; for example increasing from $n_h = 4096$ to $n_h = 8192$ yields only a 4.7\% relative improvement with respect to the extrapolated ground state.  

At large $n_h$, increasing $n_h$ is not only less effective, but also increasingly expensive. A fixed depth multi-layer MLP-NNBF scales as $\mathcal{O}(n_o n_h + n_h^2 +  \alpha n_h n_o^2+(\alpha n_o)^3)$ per step where $n_o$ is the number of orbitals and $\alpha$ is the filling fraction. For our system, $n_o=L=128$ and $\alpha=0.4375$.  
For small $n_h$, the cost of evaluating the determinant, $(\alpha n_o)^3$, dominates and the dependence on $n_h$ is primarily linear through the $n_hn_o$ term. This linear scaling is reflected in the GPU runtime in Fig.~\ref{fig:NeuronEnhancements}.  Once $n_h > \alpha n_o^2$, the $n_h^2$ term becomes dominant, and the computational cost grows quadratically with $n_h$. For our parameters, this crossover happens at $n_h=7168$ which interestingly coincides with the point where the energy gains also begin to saturate. Thus, beyond a certain network width, the benefit of increasing $n_h$ becomes marginal due to both diminishing variational improvement and rapidly increasing computational cost. This motivates the development of alternative strategies to enhance the NNBF ansatz.

\noindent \textbf{\textit{Pffafian NNBF:}} One natural change is to modify the underlying mean-field away from the Slater Determinant to BDG-NNBF \cite{OriginalNNBF} or pfaffians \cite{PP_RBM,HFPS,Pfaff_ex1,Pfaff_ex2,Pfaff_ex3}.  Here we consider a direct pfaffian NNBF (PNNBF) which we define as 
\begin{equation}
    \psi_{PNNBF}(\mathbf{n}) = \text{pf}\Big(\left(\mathbf{U}(\mathbf{n})\mathbf{J}(\mathbf{n})\mathbf{U}(\mathbf{n})^T \right)[\mathbf{n},\mathbf{n}]\Big)
\end{equation}

where $\mathbf{U}(\mathbf{n})$ and $\mathbf{J}(\mathbf{n})$ are both $L \times L$ matrices output by the MLP (e.g. the last layer of the network has $2L^2$ nodes) and $[\mathbf{n},\mathbf{n}]$ denotes selecting both the rows and columns corresponding to the occupied sites.  

This formulation strictly generalizes the standard NNBF ansatz: if $\mathbf{J}(\mathbf{n})$ is initialized such that $\text{pf}(\mathbf{J})=1$ (via output biases) and $\mathbf{U}(\mathbf{n})$ is taken to be the $N\times L$ NNBF matrix output padded with zeros to size $L\times L$, the original determinant based NNBF is recovered.

We study PNNBF with $n_h=2048$, initializing the parameters by transfer learning from a trained NNBF with the same width and $J(n)$ initialized with 1's and -1's on the super-diagonal and sub-diagonal respectively. As shown in Fig.~\ref{fig:NeuronEnhancements}, the resulting energy improves only 11\% relative to the NNBF state with the same number of neurons. In contrast, a determinant based NNBF with four times the neurons achieves slightly lower energy than the PNNBF at comparable cost. Training PNNBF from scratch yields an energy consistent with the transfer-learned result but requires approximately 20\% more GPU time. 

While the PNNBF has the same formal scaling as NNBF, we find that its optimization is significantly slower in practice.  The minimal improvement in energy is consistent with previous results on the BDG-NNBF \cite{OriginalNNBF} which found the improved mean field reference gives an advantage at fixed $n_h$ but this benefit diminishes as $n_h$ increases.

\noindent \textbf{\textit{Optimization-free multi-determinants:}} An alternative approach to improving NNBF is to introduce multiple determinants \cite{FermiNet, PauliNet}.  We consider two methods of generating multi-determinant expansions without introducing significant additional variational parameters: symmetry projection and single-step Lanczos.  We start by considering how to use these approaches with minimal additional optimization.  Later in the manuscript we will see that optimizing with a symmetrized multi-determinant expansion gains additional significant energy at some cost in time. 

While symmetries can be spontaneously broken in the thermodynamic limit, on a finite lattice the ground state must respect the symmetry group $G$ of the Hamiltonian. There are several ways to enforce symmetry in a wavefunction. For Abelian symmetries, one can choose a representative configuration for each equivalence class, $\{g\mathbf{n}: \forall g \in G\}$, and fix the magnitude of the amplitude of all configurations in that class to that of the representative, with phases set to project onto the desired symmetry sector \cite{SymCarleo}. Such constructions will often have higher energy because forcing different configurations to have the same amplitudes is an additional constraint on the variational freedom. We see this empirically (see Appendix \ref{sec:RepresentativeSymmetry}). Alternatively, equivariant convolution neural networks can encode symmetries into the mean-field orbitals producing a symmetrized wave-function with only a single network evaluation (although sometimes multiple determinant evaluations) \cite{Di_CNN, symHFDS, HFPS}. In either case, symmetry is enforced by constraining the wave-function which reduces expressivity. In contrast, our goal is to maximize representation ability while keeping the cost constrained. 

Here, we enforce symmetrization as 
\begin{equation}
    \Psi_{sym}(\mathbf{n})= \sum_g \Pi(\mathbf{n},g)\Psi(g\mathbf{n}),
    \label{eq:Symmetry}
\end{equation}
where $g \in G$ and $\Pi(\mathbf{n},g) = \pm 1$ is an additional sign coming from fermion permutations. This generates a large multi-determinant expansion in the trivial irrep, though other irreps are straightforward to implement. This requires a number of evaluations of both the network and the determinant equal to the number of symmetries, which scales linearly with system size due to translations. The $4\times16$ lattice has 4 point group symmetries, 64 translations, and an additional 2 symmetries from spin resulting in a multi-determinant expansion of 512 determinants. As shown in Fig.~\ref{fig:NeuronEnhancements}, this projection gives a consistent decrease in energy of approximately $-0.002$ (equivalent to an improvement between $27\%$ and $34\%$) for larger neuron number ($n_h>128$). Despite the large multi-determinant expansion, the optimization cost is unchanged, as the parameters are optimized only on the single-determinant ansatz. 

A second approach to generating a nearly optimization-free multi-determinant expansion is to apply a single Lanczos step. Given a trained NNBF state $\ket{\psi}$ we consider $(1+\alpha H) \ket{\psi}$ and optimize only the scalar coefficient $\alpha$. Rather then computing $\langle H^3 \rangle$ directly from sampling from $|\psi|^2$, we follow the iterative sampling method of Ref.~\cite{VMC-Lanczos}, which samples from $(1+\alpha H) \ket{\psi}$ with the current optimal $\alpha$ (see Appendix \ref{sec:lanczos}). Using this method, it takes around 10 GPU hours to determine $\alpha$, which is negligible on top of the cost of the MLP optimization. The Lanczos step generates determinants connected to $\ket{\psi}$ via the Hamiltonian. Since each spin can hop to up to 4 neighbors, the resulting multi-determinant expansion can have up to $4N$ additional determinants. For  our $\delta =1/8$ model this results in 225 determinants, fewer than in the symmetry projection. As shown in Fig.~\ref{fig:NeuronEnhancements}, the Lanczos steps performs similarly to the symmetry projection at all neurons, but gets slightly lower energies, particularly at larger $n_h$ where it provides an additional between $26\%$ and $50\%$ improvement in the energy.

These simple enhancements already achieve state-of-the-art-energies. Symmetry projection alone reaches energies comparable to the best previous results \cite{HFPS} while the multi-determinant expansion generated by a single Lanczos step on top of NNBF with either $n_h=4096$ or $n_h=8192$ gives lower energies than all previous reported values.  The NNBF training time is significantly faster ($\approx 200$ GPU hours) compared to previously reported HFPS training times  ($\approx 1500$ GPU hours \cite{HFPS}). This demonstrates that symmetry projection and Lanczos corrections provide an efficient and highly effective methodology for improving NNBF.

\noindent \textbf{\textit{Optimizing multi-determinants:}} In the previous section, we optimize the MLP-NNBF ansatz first and then project it into a multi-determinant expansion. Instead, one can optimize within the multi-determinant expansion after projection \cite{GustavoSym}.  Here we optimize the energy of Eq.~\ref{eq:Symmetry} using the parameters of a single MLP using different sets of symmetries including the 4 point group symmetries;  32 symmetries by adding in spin inversion and half translation symmetries; and 512 symmetries by including all translations. After each optimization, we also evaluate the energy after projecting onto the full 512 symmetries.  

This procedure systematically improves the ansatz and gets energies substantially lower than other enhancements. As shown in Fig.~\ref{fig:SymmetryOptimization}, for the $n_h = 2048$ model, optimization with four symmetries improves the energy by 29\% (43\% after full projection), optimization with the 32 symmetries improves the energy by 54\% (67\% after full projection), and optimization with all 512 symmetries improves the energy by 75\% relative to the single MLP ansatz with same $n_h$.  These additional optimizations all use the single-determinant NNBF as a starting point and, in total, take approximately 200, 800, and 3000 GPU hours respectively. Interestingly, as can be seen in Fig.~\ref{fig:SymmetryOptimization}, the optimization curve of the fully symmetrized state is almost always lower in energy than any other state for a given number of GPU hours. This suggests that for a fixed GPU budget an effective strategy is to simply optimize the fully symmetrized state until the target number of hours. 

We now directly compare energies for ansatzes that are evaluating an equal number of determinants. We find that our ansatz with four determinants generated by optimizing four symmetries has an energy of -0.76340(2) which is slightly lower than Transformer-BF \cite{TransformerBF} energy of $-0.76298$ which also uses four determinants. This suggest that either our $n_h=2048$ MLP has more representation ability than the Transformer-BF or that the transformer optimization is stuck in a local minima.  No explicit time is reported in Ref.~\cite{TransformerBF}.  

The HFPS \cite{HFPS} uses a CNN to augment a pfaffian ansatz using the hidden fermion methodology. The translation equivariance of the CNN architecture and a choice of a sub-lattice symmetric pfaffian ansatz enforces symmetrization without multiple network or pfaffian evaluations. However, to restore the full symmetric invariance, there is a sum over 32 pfaffians to restore all translational, point group and spin symmetries. The HFPS reaches an energy of $-0.76413$ which is somewhat higher then then $-0.76486(2)$ of our symmetry optimization with 32 determinants.  Additionally, our result can be further projected at essentially no cost to the fully symmetric state to reach an energy of $-0.76560(1)$, while the HFPS can not be projected further. We suspect that this improvement in energy is largely due to the fact that the hidden fermion methodology is essentially a low rank NNBF update and therefore has lower representation ability compared to NNBF \cite{ZejunHFDS}.  The optimization of our 32 determinant MLP-NNBF is faster by approximately a factor of 2 compared with the HFPS.

One could further relax constraints and optimize all the determinants separately. We find that the energy landscape with multi-determinant expansions is challenging, as when we initialize a model with 4 independent determinants starting from the symmetric point, we find no improvement in energy (and actually an increase in energy at fixed learning rate suggesting a need for more adaptive optimization). This suggests that by constraining the ansatz we are helping simplify the energy landscape.

\noindent \textbf{\textit{Fixed Node Diffusion Monte Carlo:}}  So far we have primarily considered variational wave-functions which can be compactly represented. Alternatively, we can use quantum Monte Carlo which represents the wave-function with an exponential number of determinants stochastically.  Because of the sign problem, we use the fixed-node diffusion Monte Carlo (FNDMC) starting with the single determinant ansatz. FNDMC takes approximately 20 additional GPU hours.  As seen in Fig.~\ref{fig:NeuronEnhancements}, the FNDMC decrease in energy has a much larger spread than either the symmetry projection or the Lanczos step spanning from as little as $7\%$ improvement to as high as a $72\%$ improvement from the NNBF at the same $n_h$.  Interestingly, while the standard lore is that FNDMC will generically do better when applied to anzats with variationally smaller energies, we do not find that to be the case here. While the energy of the NNBF decreases monotonically with $n_h$, the energy of FNDMC vacillates with $n_h$. This suggests that at these small energy scales it would be better to directly optimize the FNDMC energy.

\begin{figure*}[htbp]
    \centering
    \includegraphics[width=.98\textwidth]{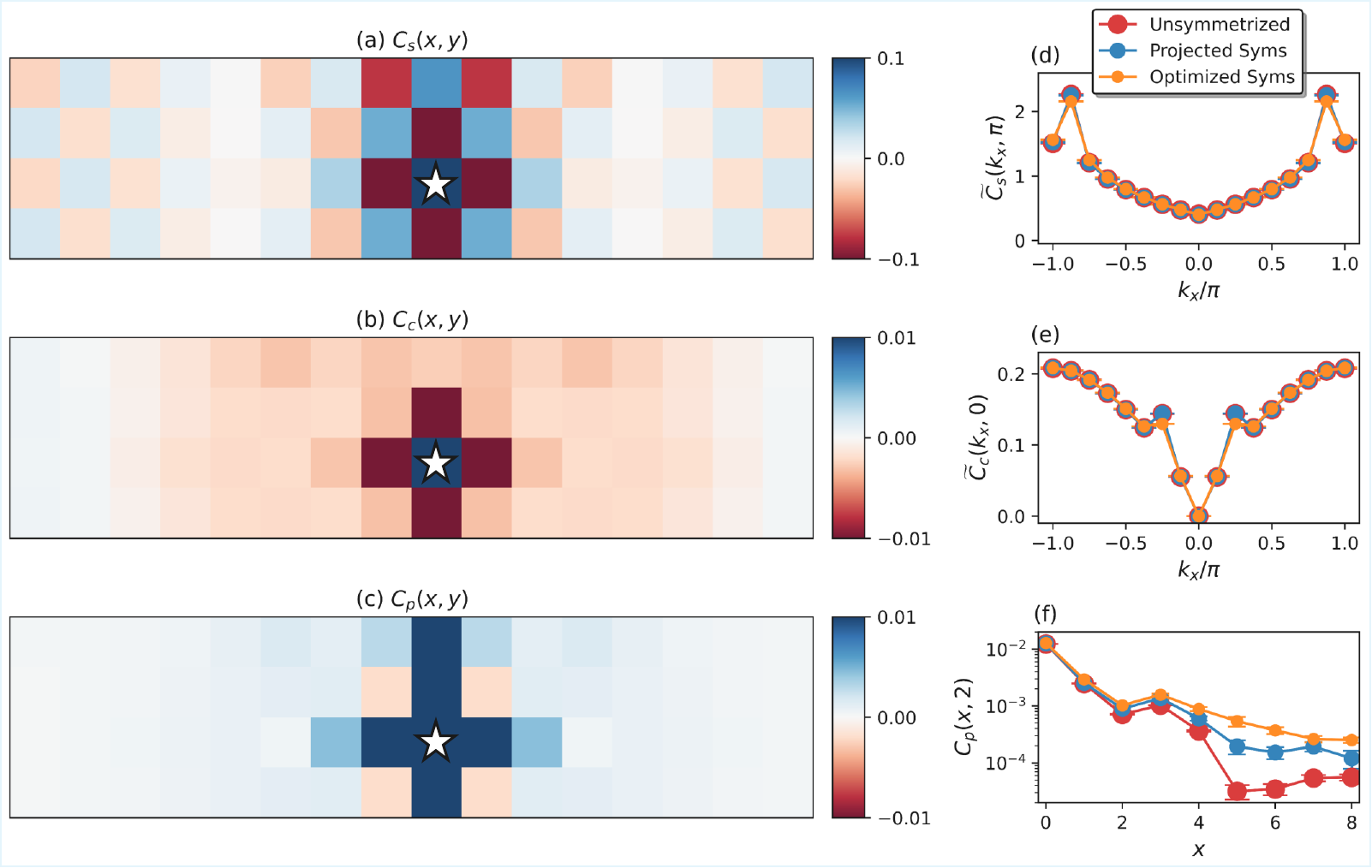}
    \caption{Observables on the $4\times 16$ lattice. (a-c) The correlation functions $C_s$, $C_c$, and $C_p$ in real space for the best variational model, $n_h=2048$ with optimized symmetries. To show weaker correlation, the values are truncated such that $\abs{C_s} \leq 0.1$ and $\abs{C_c}, \abs{C_p} \leq 0.01$. The central site is marked with a star. (d,e) The spin and charge structure factors of $n_h=2048$ models with no symmetries, projected symmetries, and optimized symmetries. Peaks appear at $(7\pi/8, \pi)$  and $(\pi/4, 0)$ respectively. (f) The pair-pair correlation $C_p(x,2)$ on the same models in $(d,e)$. Similar benchmarks for HFPS are shown in Ref.~\cite{HFPS}.}
    \label{fig:ObservableComparison}
\end{figure*}

\subsection{\label{subsec:Observables}Observables}

In this regime, the Hubbard model is believed to have stripes with anti-ferromagnetic (AFM) order separated by one-dimensional hole-rich domain walls over which the polarization of the AFM order flips \cite{ODLRO_ex4,ODLRO_ex1,ODLRO_ex2,ODLRO_ex3,HFPS,TransformerBF}.   We compute observables for our different variational wave-functions including the spin-spin correlations
\begin{equation}
    C_s(\mathbf{r}) = \frac{1}{\ell}\sum_\mathbf{i}\langle\mathbf{\hat{S}}_{\mathbf{i}+\mathbf{r}} \cdot \mathbf{\hat{S}}_\mathbf{i}\rangle - \left(\frac{1}{\ell}\sum_\mathbf{i} \langle\mathbf{\hat{S}}_\mathbf{i}\rangle\right)^2
\end{equation}
and the charge-charge correlations 
\begin{equation}
    C_c(\mathbf{r}) = \frac{1}{\ell}\sum_\mathbf{i}\langle {\hat{n}}_{\mathbf{i}+\mathbf{r}} \cdot{\hat{n}}_\mathbf{i}\rangle -  \left(\frac{1}{\ell}\sum_\mathbf{i} \langle \hat{n}_\mathbf{i}\rangle\right)^2
\end{equation}
where the sum is over the physical site locations and $\ell \equiv L/2$ is the total number of physical sites.  
In Fig.~\ref{fig:ObservableComparison}(a,b), we show these observables for our lowest energy variational wave-function.  As anticipated, there are two domain walls across which the AFM sub-lattice polarization flips. One can see from the charge-charge correlation, these domain walls are hole-doped.  
We look at the spin ($\alpha=s$) and charge ($\alpha=c)$ Fourier transforms 
\begin{equation}
    \widetilde{C}_\alpha(\mathbf{k}) = \sum_{\mathbf{r}}C_\alpha(\mathbf{r})e^{-i\mathbf{k}\cdot\mathbf{r}}
\end{equation}
in Fig.~\ref{fig:ObservableComparison}(d,e) for a variety of different variational wavefunctions. Looking at slices in the $k$-space structure factors at $\mathbf{k} = (k_x, \pi)$ for the spin and $\mathbf{k} = (k_x, 0)$ for the charge, we find the shown variational wave-functions look similar and have respective peaks at $\mathbf{k} = (7\pi/8, \pi)$ and $\mathbf{k} = (\pi/4, 0)$. This demonstrates that even higher energy NNBF variational wave-functions capture the spin density wave with period 16 and charge density wave with period 8 successfully. Interestingly, if we go to the $n_h=128$ state which is much higher in variational energy, there is a non-trivial change in the width of the stripes and therefore respective structure factors (see Fig.~\ref{fig:n128obs}).

The doped square lattice Hubbard model can also have a tendency towards d-wave pairing. We compute the pair-pair correlation function 
\begin{equation}
    C_p(\mathbf{r}) = \frac{1}{\ell}\sum_\mathbf{i}\langle \hat{\Delta}^\dagger(\mathbf{i}+\mathbf{r}) \hat{\Delta}(\mathbf{i})\rangle,
\end{equation}
for d-wave symmetrized pairs
\begin{equation}
    \hat{\Delta}(\mathbf{r}) = \frac{1}{4}\sum_\mathbf{\delta}\frac{1}{\sqrt{2}}\text{sign}(\delta)(\hat{c}_{\mathbf{r},\uparrow}\hat{c}_{\mathbf{r}+\delta,\downarrow}-\hat{c}_{\mathbf{r},\downarrow}\hat{c}_{\mathbf{r}+\delta,\uparrow}).
\end{equation}

As shown in Fig.~\ref{fig:ObservableComparison}(f), we find a non-trivial short-range d-wave pairing.  A strong short-range d-wave pairing was also seen in Ref. ~\onlinecite{HFPS} but required a pfaffian and optimization using initialized (but eventually removed) pinning fields to find.  Interestingly, the large $x$ pairing function increases as the wave-function quality improves and there are hints that our pairing function is starting to saturate at the largest $x$. 

\subsection{\label{subsec:4x8Lattice}$4 \times 8$ Lattice}

\begin{figure*}[htbp]
    \centering
    \includegraphics[width=.98\textwidth]{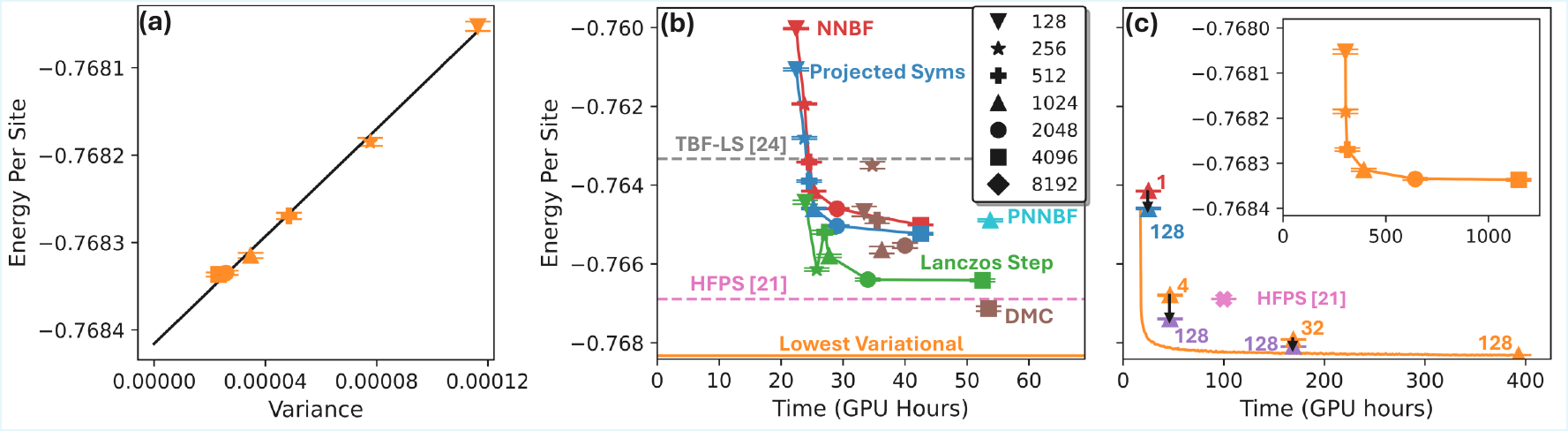}
    \caption{Energies on the $4\times 8$ lattice. Symbols denote $n_h$. (a) Energy vs variance of the optimized symmetries at various $n_h$ with variance extrapolated best-fit line which is used as the bottom for the y-axis in (b) and (c).
    (b) Energy vs GPU time for different methods which optimize only the single mean-field as well as variational energies from other works (dotted lines) and our lowest variational energy (solid orange).  
    (c) Energy vs GPU time for optimized symmetries (orange) and projected optimized  symmetries (purple) at  $n_h=1024$ with numbers indicating the number of symmetries.  The orange optimization curve is for the projected optimized symmetry with 128 symmetries.  Inset contains projected optimized symmetries with 128 symmetries at various $n_h$.}
    \label{fig:4x8}
\end{figure*}

In addition to the $4\times 16$ lattice, we benchmark these enhancements on a $4 \times 8$ lattice and find generically similar results as shown in Fig.~\ref{fig:4x8}. 
Most strikingly, the lowest variational energy, produced by optimizing the fully symmetrized state over 128 determinants, reaches an energy of $-0.768337(2)$ which has a relative error of $1.0\times 10^{-4}$ compared to the variance extrapolated ``ground truth'' energy of $-0.768416$.  This is an order-of-magnitude improvement compared to the previous best variational energies on this model\cite{HFPS}.

\section{\label{sec:discussion}Discussion}
In this work, we have investigated several strategies for improving the NNBF ansatz.  Using these strategies, we find the ansatz with the lowest reported variational energies for the $U=8$ Hubbard model with $\delta =1/8$ hole doping in periodic boundary conditions for both the $4 \times 16$ and $4 \times 8$ lattices.  While increasing the width of an MLP will decrease the energy, in practice we find that at large enough neurons there is diminishing returns in this process and a multi-determinant approach becomes preferable. Enhancing an optimized NNBF ansatz with a single Lanczos step, diffusion Monte Carlo, or symmetry projection yields substantial energy improvements at almost no cost in time yielding results for the $16 \times 4$ lattice that are both lower in energy and faster (where reported) than previous techniques. By further optimizing after projection, one gets significant additional gains in energies.  This is the case even when optimizing with only a small number of such determinants (e.g. 32) and then projecting afterwards.  That said, interestingly, at any fixed amount of GPU time, it is nearly optimal to just optimize the fully symmetrized state for that number of GPU hours. The uniformity of symmetrized NNBF wave-functions permitted accurate variance extrapolated energies. We also considered various observables for these systems finding spin and charge density waves that are consistent with previous results and finding that the d-wave pair correlation functions have strong local correlations and show hints of saturation at large $x$.
A natural next step is to apply the symmetrized and post-processed NNBF framework to additional challenging and physically interesting systems. 

\begin{acknowledgments}

We acknowledge the helpful conversations with Z. Liu, A. Liu, and J. Hahm. The NQS simulations are implemented using internally developed code with JAX \cite{JAX}, Flax \cite{flax}, and Optax \cite{optax}. This material is based upon work supported by the U.S. Department of Energy, Office of Science, National
Quantum Information Science Research Centers.
This research used both the DeltaAI advanced computing and data resource, which is supported by the National Science Foundation (award OAC 2320345) and the State of Illinois, and the Delta advanced computing and data resource which is supported by the National Science Foundation (award OAC 2005572) and the State of Illinois. Delta and DeltaAI are joint efforts of the University of Illinois Urbana-Champaign and its National Center for Supercomputing Applications.
\end{acknowledgments}

\bibliography{bibliography}

\appendix
\renewcommand{\thefigure}{A\arabic{figure}}
\setcounter{figure}{0}

\section{\label{sec:NeuronOptimization}Neuron Optimization}

\begin{figure}[htbp]
    \centering
    \includegraphics[width=0.48\textwidth]{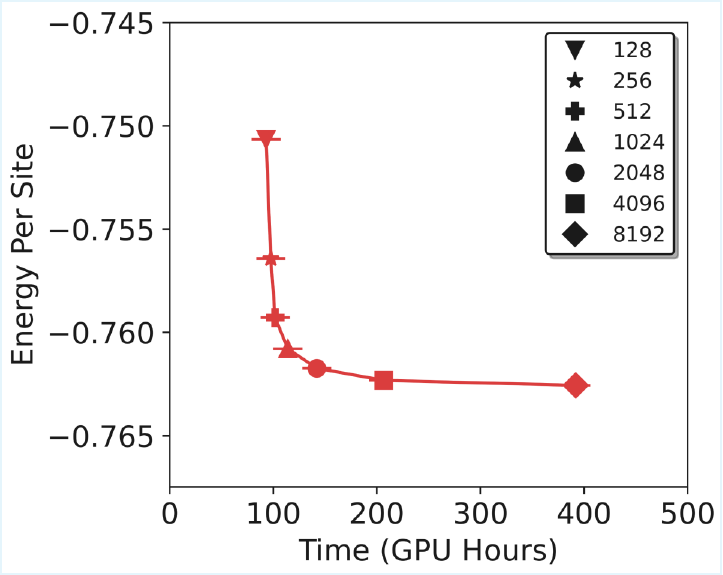}
        \caption{Variational energy as a function of GPU training time at various $n_h$ demonstrating the trade-off between model expressiveness and computational cost. Symbols denote $n_h$.}
    \label{fig:BaseNeurons}
\end{figure}

In Fig.~\ref{fig:BaseNeurons} we present the time taken to perform the optimization of the base NNBF ansatz on the $4 \times 16$ lattice with varying neurons zoomed out compared to Fig.~\ref{fig:NeuronEnhancements}. As mentioned in section \ref{sec:results}, we see that the energy decreases rapidly with neurons at small $n_h$ but improves much more slowly as $n_h$ increases. At large $n_h$, the energy effectively plateaus significantly above the extrapolated ground state.

\section{\label{sec:OptimizationDetails}Optimization Details}

Unless otherwise specified, the total number of markov chains is fixed at $N_s = 1024$ and the learning rate is fixed at $3\times 10^{-4}$. We first train the single particle orbitals without backflow for $10^5$ VMC iterations. We then multiply all single particle orbitals by a factor of $10^3$, set them to the bias of the last layer of the MLP, and randomly initialize the weights to form the initial MLP-NNBF ansatz. The training for the data in Fig.~\ref{fig:BaseNeurons} took $3\times 10^6$ VMC iterations, followed by $5 \times 10^5$ VMC iterations with an exponential decay learning rate schedule tuning the learning rate to $10^{-7}$.

The training for the model with 4 symmetries took approximately $8\times10^5$ VMC iterations, while the training for the model with 32 symmetries took approximately $5\times10^5$ VMC iterations. The training for the model with all symmetries took $1.5 \times 10^5$ VMC iterations. The training for these models all ended with an exponential decay learning rate schedule tuning the learning rate to $10^{-7}$. The total times for training these models is shown in Fig.~\ref{fig:SymmetryOptimization}.

\section{\label{sec:RepresentativeSymmetry}Representative Symmetry}

\begin{figure}[htbp]
    \centering
    \includegraphics[width=0.48\textwidth]{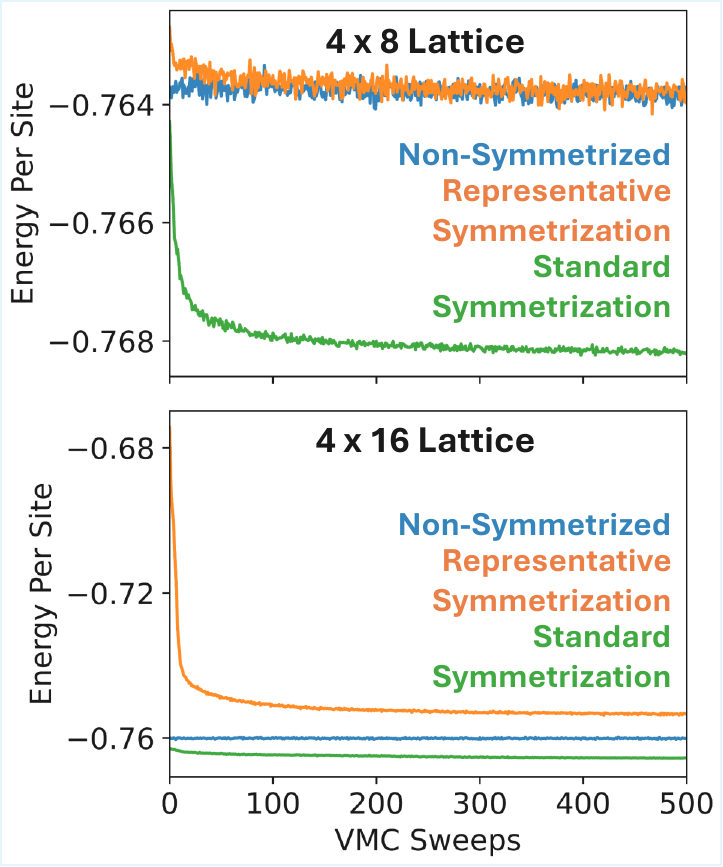}
    \caption{Energy vs VMC sweep. The representative symmetrization and standard symmetrization are started from the non-symmetrized wavefunction. We see in both the $4\times 8$ ($n_h=1024$) and $4\times 16$ ($n_h=2048$) lattices, the representative symmetrization increases the energy, while the standard symmetrization decreases it. Each VMC sweep reports the average energy from 100 VMC steps of each of the 1024 VMC chains.}
    \label{fig:RepresentativeSymmetries}
\end{figure}

As mentioned in Section~\ref{sec:methods}, there are many different ways to ensure a wave-function preserves symmetry. Here we consider symmetrization by choosing a representative configuration for each equivalence class to fix the magnitude of the amplitude for all configurations in that group and with phases then fixed by the symmetry sector. 
We see empirically in Fig.~\ref{fig:RepresentativeSymmetries} that choosing a representative and projecting to the trivial irrep initially increases the energy, and while the energy eventually comes down to the base energy in the $4 \times 8$ lattice, the energy in the $4 \times 16$ lattice ends up higher then the base model. 

\section{\label{sec:lanczos}Lanczos Step}
Lanczos steps improve the representation ability of a variational ansatz by projecting the wavefunction into the Krylov subspace of the Hamiltonian. The first Lanczos step can be found by choosing the optimal alpha to construct $\ket{\psi_{\alpha}}  = (1 + \alpha H) \ket{\psi_i}$ where $\ket{\psi_i}$ is the previously optimized wavefunction \cite{VMC-Lanczos, TBF-Lanczos, Lanczos-Example}. Calculating the optimal $\alpha^*$ requires minimizing the energy
\begin{align}
    E(\alpha) = \frac{\bra{\psi_i}(1+\alpha H) H (1+\alpha H) \ket{\psi_i}}{\bra{\psi_i}(1+\alpha H)^2
    \ket{\psi_i}} \label{eq:EofAlpha}
\end{align}
of the new ansatz. The standard method for calculating a single Lanczos step requires computing the powers of the Hamiltonian 
\begin{equation}
    h_n = \frac{\bra{\psi}H^n\ket{\psi}}{\braket{\psi|\psi}} \label{eq:h_n}
\end{equation}
up to $n=3$, from which one can analytically solve for $\alpha^*$. 

In practice we implement the algorithm in Appendix C of Ref. \cite{VMC-Lanczos}. We start by guessing a value of $\alpha$ and computing $E(\alpha)$ from Eq. \ref{eq:EofAlpha} and 
\begin{align}
    \chi &= \frac{\bra{\psi_\alpha}(1+\alpha H)^{-1}\ket{\psi_\alpha}}{\langle{\psi_\alpha}
    \ket{\psi_\alpha}} \nonumber \\
    &= \frac{\bra{\psi_\alpha}\ket{\psi_i}}{\langle{\psi_\alpha}(1+\alpha H)
    \ket{\psi_i}} \label{eq:Chi} \\
    &= \langle [1+\alpha E_i(\mathbf{n})]^{-1} \rangle_{\abs{\psi_\alpha}^2} \nonumber
\end{align}
where $E_i(\mathbf{n}) = \bra{\psi_i}H\ket{\mathbf{n}}/\langle{\psi_i}\ket{\mathbf{n}}$ is the local energy of the initial wavefunction. From this we can obtain 
\begin{equation}
    h_2 = [(\chi^{-1}-2)(1+\alpha h_1)+1]/\alpha^2,    
\end{equation}
and 
\begin{equation}
    h_3 = [E(\alpha)(1+2\alpha h_1 +\alpha^2 h_2) -h_1-2\alpha h_2]/\alpha^2,
\end{equation}
which can be used to analytically minimize $E(\alpha)$ for the optimal $\alpha^*$. Note that $h_3$ is found by sampling an energy expectation value, which is statistically more accurate compared to the direct determination of $h_3$ from equation \ref{eq:h_n}.

The analytical minimization of $E(\alpha)$ is statistically most accurate as $\alpha$ gets closer to $\alpha^*$, so we typically iterate quickly by using a small number of samples to estimate $E(\alpha)$ and $\chi$ until we are close to $\alpha^*$. In practice we begin with $\alpha$ near zero, and estimate $h_n$ with $2^{16}$ samples. We usually find good convergence within 4 iterations, after which we iterate two more times with $2^{18}$ samples to further lower statistical error. The $\alpha^*$ for the four best wavefunctions in Fig.~\ref{fig:NeuronEnhancements} are listed in Table \ref{tab:Lanczos}.

\vspace{1em}
\begin{table}[htbp]
\centering
\setlength{\tabcolsep}{12pt}
\renewcommand{\arraystretch}{1.3}
\begin{tabular}{@{}c c@{}}
\toprule
$n_h$ & $\alpha^*$ \\
\midrule
1024  & 0.03233 \\
2048  & 0.03875 \\
4096 & 0.02641 \\
8192 & 0.02898 \\
\bottomrule
\end{tabular}
\caption{Optimized $\alpha^*$ values for different $n_h$ values on the $4\times 16$ Hubbard model. The energies of the resulting wavefunctions $\ket{\psi_{\alpha^*}}  = (1 + \alpha^* H) \ket{\psi_i}$ are shown in Fig.~\ref{fig:NeuronEnhancements}. }
\label{tab:Lanczos}
\end{table}

\section{\label{sec:HighEnergyObservables}High Energy Observables}
\begin{figure*}[htbp]
    \centering
    \includegraphics[width=0.98\textwidth]{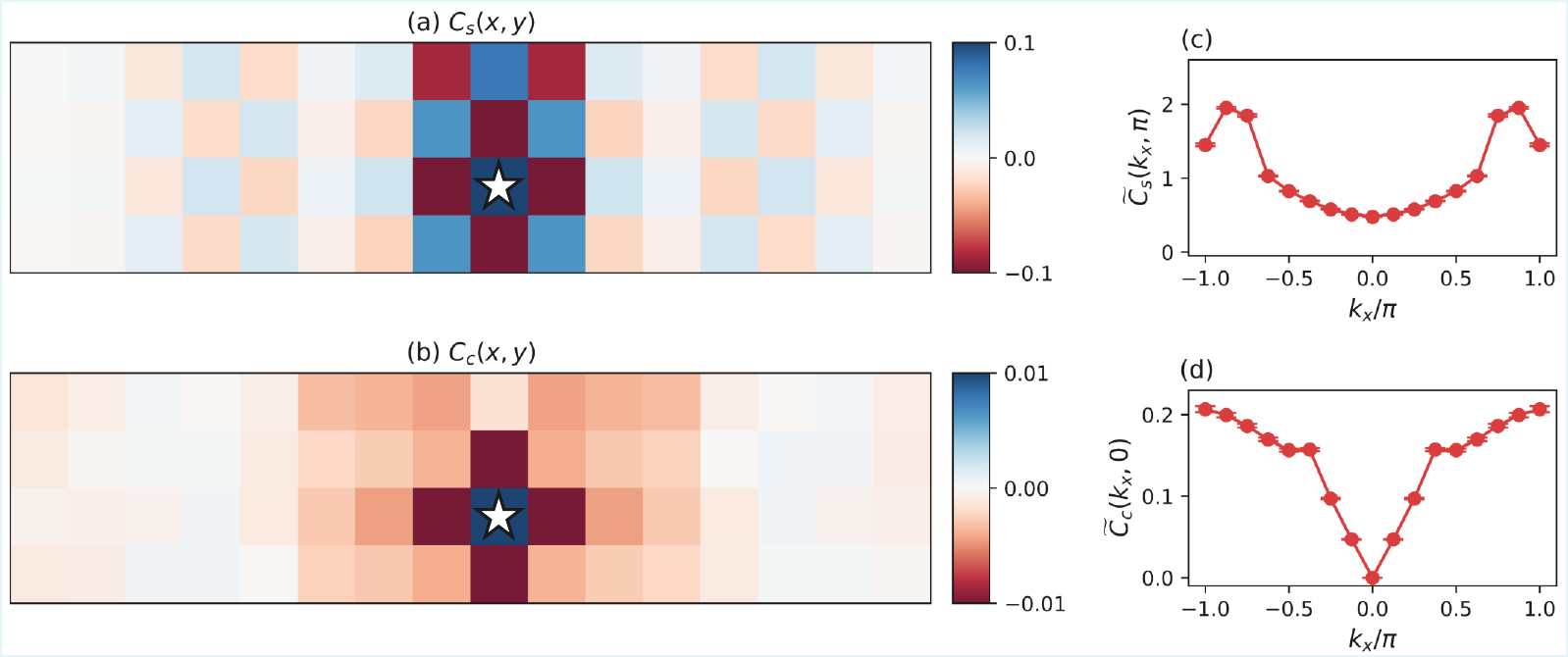}
    \caption{Spin and charge observables for base NNBF with $n_h = 128$. This ansatz has a high energy per site of $-0.75066(2)$. (a-b) The correlation functions $C_s$ and $C_c$ in real space. The values are truncated such that $\abs{C_s} \leq 0.1$ and $\abs{C_c} \leq 0.01$ to show weaker correlations. The central site is marked with a star. (d-e) The spin and charge structure factors showing cuts at $\widetilde{C}_s(k_x,\pi)$ and $\widetilde{C}_c(k_x,0)$. Peaks seem to be $(7\pi/16, \pi)$  and $(3\pi/8, 0)$ respectively instead of at the correct values of $(7\pi/8, \pi)$  and $(\pi/4, 0)$ found for more accurate states in Fig.~\ref{subsec:Observables}.}
    \label{fig:n128obs}
\end{figure*}

As mentioned in Sec. \ref{subsec:Observables}, most of the wavefunctions, even prior to enhancement are able to correctly capture the spin and charge density waves. However, the highest energy wavefunction we study, $n_h=128$ with energy per site of $-0.75066(2)$, does not capture the density waves correctly. As shown in Fig.~\ref{fig:n128obs} (c,d), we see a spin density wave with a peak at approximately $\mathbf{k}=(7\pi/16,\pi)$ and a charge density wave with peak at $\mathbf{k}=(3\pi/8,\pi)$. This suggests that a large enough number of neurons is required to capture these structures accurately.

\end{document}